\documentclass[aps,prl,twocolumn,preprintnumbers,amsmath,amssymb,superscriptaddress,10pt]{revtex4-1}
\usepackage{graphicx}
\usepackage{bm}
\usepackage[dvipsnames]{xcolor}
\usepackage{color}
\usepackage{braket}
\usepackage{amsmath}
\usepackage{amssymb}
\usepackage{amsfonts}
\usepackage[T2A]{fontenc}
\usepackage{enumitem}
\usepackage[colorlinks]{hyperref}
\usepackage[cp1251]{inputenc}

\begin{document}

\title{Gigantic spin-noise gain enables magnetic resonance spectroscopy of impurity crystals}

\author{A.~N. Kamenskii}
\affiliation{Experimentelle Physik 2, Technische Universit\"at Dortmund, 44221 Dortmund, Germany}

\author{A. Greilich$^*$}
\affiliation{Experimentelle Physik 2, Technische Universit\"at Dortmund, 44221 Dortmund, Germany}

\author{I.~I. Ryzhov}
\affiliation{Photonics Department, St.\,Petersburg State University, Peterhof, 198504 St.\,Petersburg, Russia}
\affiliation{Spin Optics Laboratory, St.\,Petersburg State University, Peterhof, 198504 St.\,Petersburg, Russia}

\author{G.~G. Kozlov}
\affiliation{Spin Optics Laboratory, St.\,Petersburg State University, Peterhof, 198504 St.\,Petersburg, Russia}

\author{M. Bayer}
\affiliation{Experimentelle Physik 2, Technische Universit\"at Dortmund, 44221 Dortmund, Germany}
\affiliation{Ioffe Institute, Russian Academy of Sciences, 194021 St.\,Petersburg, Russia}

\author{V.~S. Zapasskii}
\affiliation{Spin Optics Laboratory, St.\,Petersburg State University, Peterhof, 198504 St.\,Petersburg, Russia}

\maketitle

\textbf{
	Spin noise spectroscopy is a method of magnetic resonance~\cite{AZ81} widely used, nowadays, in atomic and semiconductor research~\cite{Muller,vzap,rise,glaz}.  Classical objects of the EPR spectroscopy  -- dielectrics with paramagnetic impurities -- seemed to be unsuitable for this technique because of large widths of allowed optical transitions and, therefore, low specific Faraday rotation (FR).  We show, however, that the FR noise detected at the wavelength of a weak optical transition (with low regular FR) may increase by many orders of magnitude as its homogeneous width decreases. This {\it spin-noise gain effect},  numerically described by the ratio of the inhomogeneous linewidth to homogeneous, relates primarily to forbidden intraconfigurational transitions of impurity ions with unfilled inner electronic shells. Specifically, for the f-f transitions of rare-earth ions in crystals, this factor may reach $\sim 10^8$.  In this paper, we report on the first  successful application of spin noise spectroscopy for detecting magnetic resonance of rare-earth ions in crystals.
}

The spectroscopy of magnetic resonances can be applied to virtually all quantum objects possessing angular momentum and, therefore, is one of the key techniques of contemporary physical research.  Detection of the
magnetic resonance spectrum usually implies the observation of a response of a paramagnet (`spin-system') to an alternating (AC) magnetic field with frequency close to that of the spin precession. Since the discovery of
the magnetic resonance effect by E. Zavoisky in 1944~\cite{Zav44}, two main detection methods have been elaborated. One of them implies detecting changes of the AC field, as done in the conventional optical spectroscopy
of absorption or reflection. In the second method, the resonance is revealed by changes in the object's properties as done, e.g., in the optical pump-probe spectroscopy. However, the principles of resonance detection in
both cases are the same: the resonance is observed only through a response to the excitation of spin precession by the external AC field.

In 1981~\cite{AZ81}, it was demonstrated experimentally that observation of the magnetic resonance does not necessarily require application of the resonant AC field -- the resonance can be observed also by the excess
fluctuations of the medium's refractive index at the spin-precession frequency. In the last years, this spin noise spectroscopy (SNS) has been considerably developed~\cite{mitsui,Crooker1,Crooker2,Oestr,charge} and revealed a number of unique capabilities that were not foreseen
initially~\cite{Glazov}. Specifically, the SNS sensitivity was found to be sufficient to detect the spin noise (SN) of a single charge carrier~\cite{single1,single2}, was applied to monitor the nuclear dynamics of the
host lattice in a doped semiconductor~\cite{nuclear1,nuclear2}, was used as an instrument of high-resolution tomography~\cite{tomogr} and as a spectroscopic tool to distinguish between the homogeneous and inhomogeneous
broadening and to measure homogeneous width of optical transitions~\cite{OSN,Yang}, etc. As a result, SNS has nowadays turned into an important method of research in atomic and semiconductor physics.

It should be noticed, however, that this method has never been applied to dielectrics with paramagnetic impurities -- systems which served as primary material basis for electron paramagnetic resonance (EPR) spectroscopy and
continue to be used in state-of-the-art optics and photonics~\cite{Faraon15}. The reason for this is the low efficiency of the conversion ''magnetization -> FR'', from which one might expect also a low efficiency of the
conversion ''magnetization noise -> FR noise''. Curiously, this is by far not the case.

\begin{figure}
\centering
\includegraphics[width=0.8\linewidth]{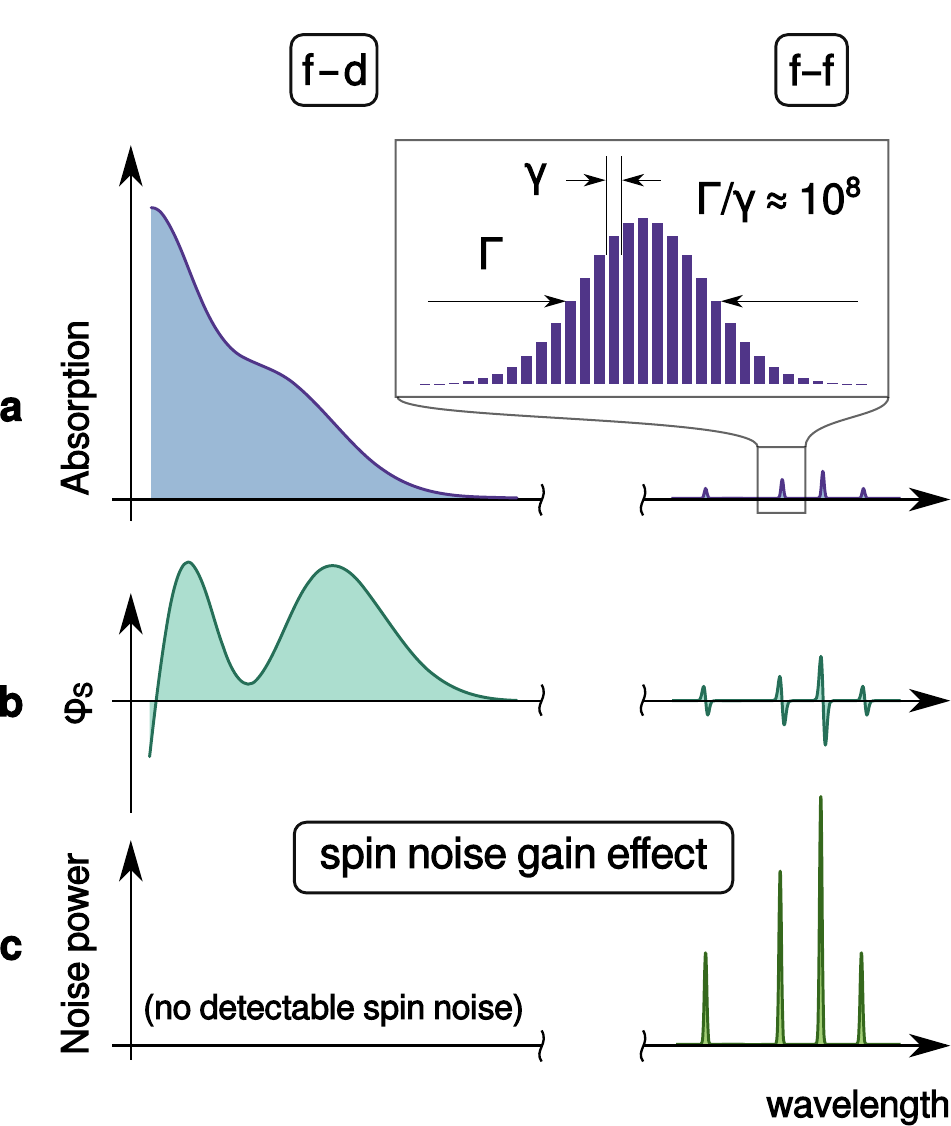}
\caption{\textbf{Schematic of spectra for rare-earth ions in crystals.} \textbf{a}, Typical spectra of absorption. \textbf{b}, Faraday rotation $\varphi_s$ and \textbf{c}, SN power. The left and right (short- and
long-wavelength) parts of the spectra show the bands of inter- and intraconfigurational transitions of the ions, respectively. The weak FR of f-f transitions may give rise to a strongly enhanced spin noise due to a large
ratio of $\Gamma/\gamma$.}
\label{fig:1}
\end{figure}

It is useful to remind here that ions with unfilled inner shells, representing the bulk of possible impurities, are important for science and applications (like Cr$^{3+}$ and Nd$^{3+}$ in ruby and neodymium lasers,
respectively, or the rare-earth (RE) ions in up-converters or quantum counters, etc.) and are typically characterized by two types of optical transitions: strong (parity-allowed) interconfigurational and weak
(parity-forbidden) intraconfigurational transitions (shown schematically in Fig.~\ref{fig:1}). The former are often characterized by a strong magneto-optical activity and therefore considered as the most promising
candidates for corresponding applications. We will restrict ourselves to the widely applied RE ions in crystals.

The applicability of SNS to crystals with paramagnetic impurities has been assessed in Ref.~\cite{FTT}, using as an example the crystal CaF$_2$ doped with divalent thulium. This ion is characterized by strong allowed
interconfigurational (4f$^n$-4f$^{n-1}$5d) transitions in the visible spectral range and is known for its among the RE ions strongest magneto-optical activity~\cite{Shen}. For this evaluation, we used the value of the FR
cross-section given in Ref.~\cite{Giri} that is numerically equal to the FR angle per unit impurity concentration and per unit length of the medium. We found that this quantity, even for such a favorable object as the
CaF$_2$:Tm$^{2+}$ crystal, is around four orders of magnitude smaller than, e.g., for n-GaAs.

For the forbidden intraconfigurational (4f - 4f) transitions, revealed for virtually  all trivalent RE ions also in the visible spectral range, the situation appears to be even worse~\cite{minerals}. The magneto-optical
activity of these transitions and the corresponding values of the FR cross-section are expected to be even smaller than for the 4f-5d transitions, so that application of SNS to these objects looks not promising at all. In
stark contrast, we will show that the FR cross section, generally, cannot be used to assess the applicability of SNS to spin-systems.\\

\textbf{The spin-noise gain effect.}
It has been shown in Ref.~\cite{OSN} that the shape of the optical spectrum (wavelength dependence) of the SN power, detected within the optical transition, is strongly affected by inhomogeneous broadening of the
transition. Here, we concentrate on the effect of {\it SN gain} revealed as a gigantic enhancement of the SN on transitions with large inhomogeneous-to-homogeneous linewidth ratios.

We consider the case when the frequency of the probe monochromatic light is within the absorption profile of the paramagnet under study ({\it resonant probing}). Let the inhomogeneous and homogeneous linewidths of the
transition be $\Gamma$ and $\gamma$, respectively. It is convenient for our reasoning to choose the wavelength of the probe beam such that the regular FR of the system is essentially different from zero (somewhere on the
flanks of the absorption line). We consider the case of total spin polarization when all spins are oriented along the light propagation direction (say, $S_i$ = +1/2 for all $i$) and the magnitude of the FR $\varphi_s$
({\it FR of saturation}) is the greatest. Let us express the angle $\varphi_s$ through the length of the sample $d$ and impurity concentration $n$ (spin density) using the notion of the FR cross-section $\sigma$
introduced in
Ref.~\cite{Giri}: $\varphi_s=\sigma n d$. The number of impurity centers within the probe beam volume $N$ is given by $N=ndA$, where $A$ is the beam cross-section.

Because of the inhomogeneous broadening of the line, the FR (being itself independent of the homogeneous width $\gamma$) is contributed only by a fraction of impurity centers with optical frequencies within an interval of
the order of the homogeneous linewidth $\gamma$. This number $N'$, is approximately equal to $N\gamma/\Gamma$, and thus evidently can drastically differ from the total number of impurity centers $N$.

Now, assuming that the total FR is the sum of contributions of the individual centers, we can write the FR per unit center $\varphi_i$ as $\varphi_i=  \varphi_s/N' \sim (\Gamma/\gamma) \varphi_s/N$. When detecting the
{\it FR noise} (rather than the standard FR), the magnitude of the contribution of each center will remain $ \varphi_s/N'$ as before, but its sign will be random: $\varphi_i= \pm (\Gamma/\gamma) \varphi_s/N$.

\begin{figure*}[ht]
\centering
\includegraphics[width=0.32\linewidth]{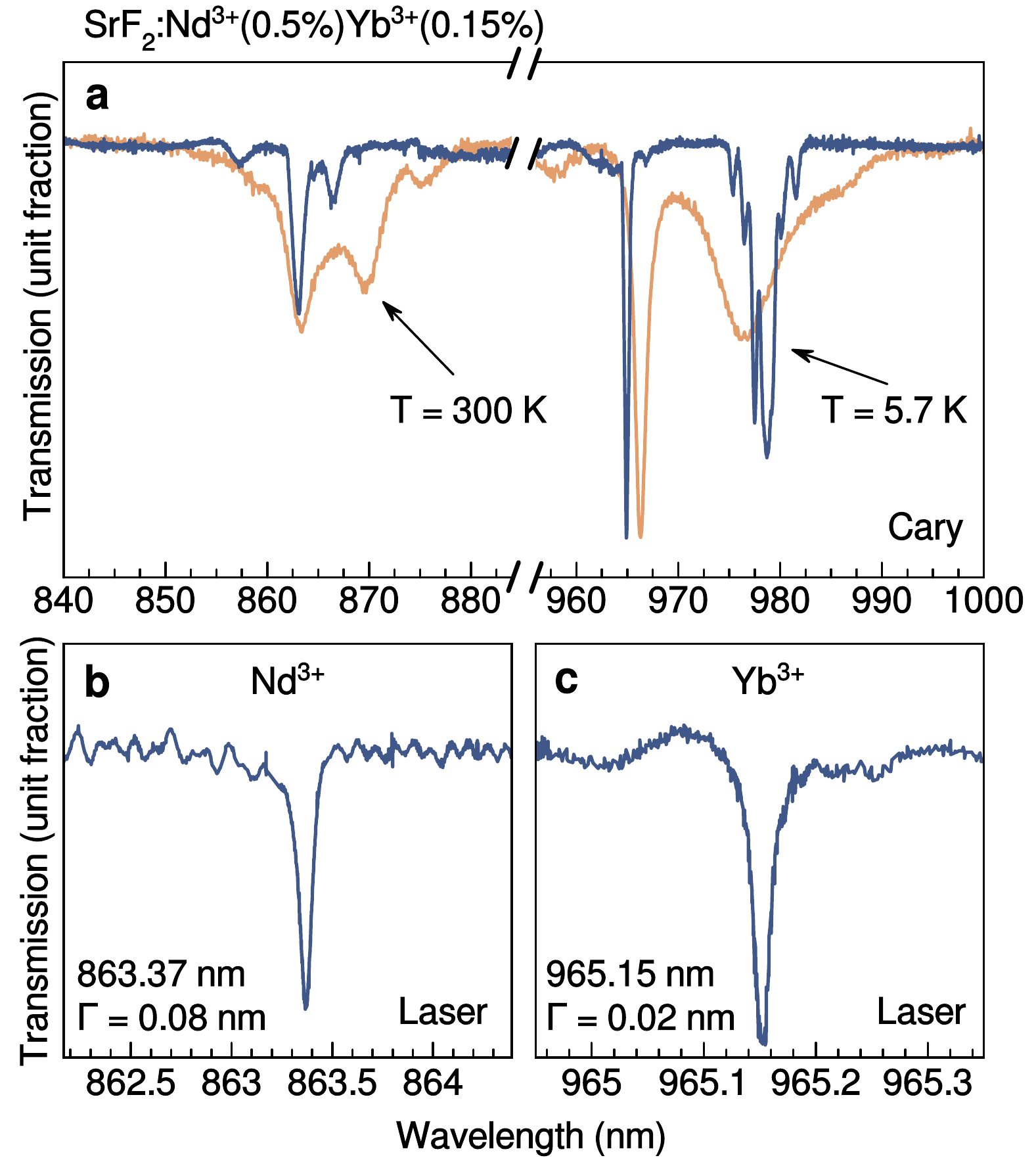}
\includegraphics[width=0.32\linewidth]{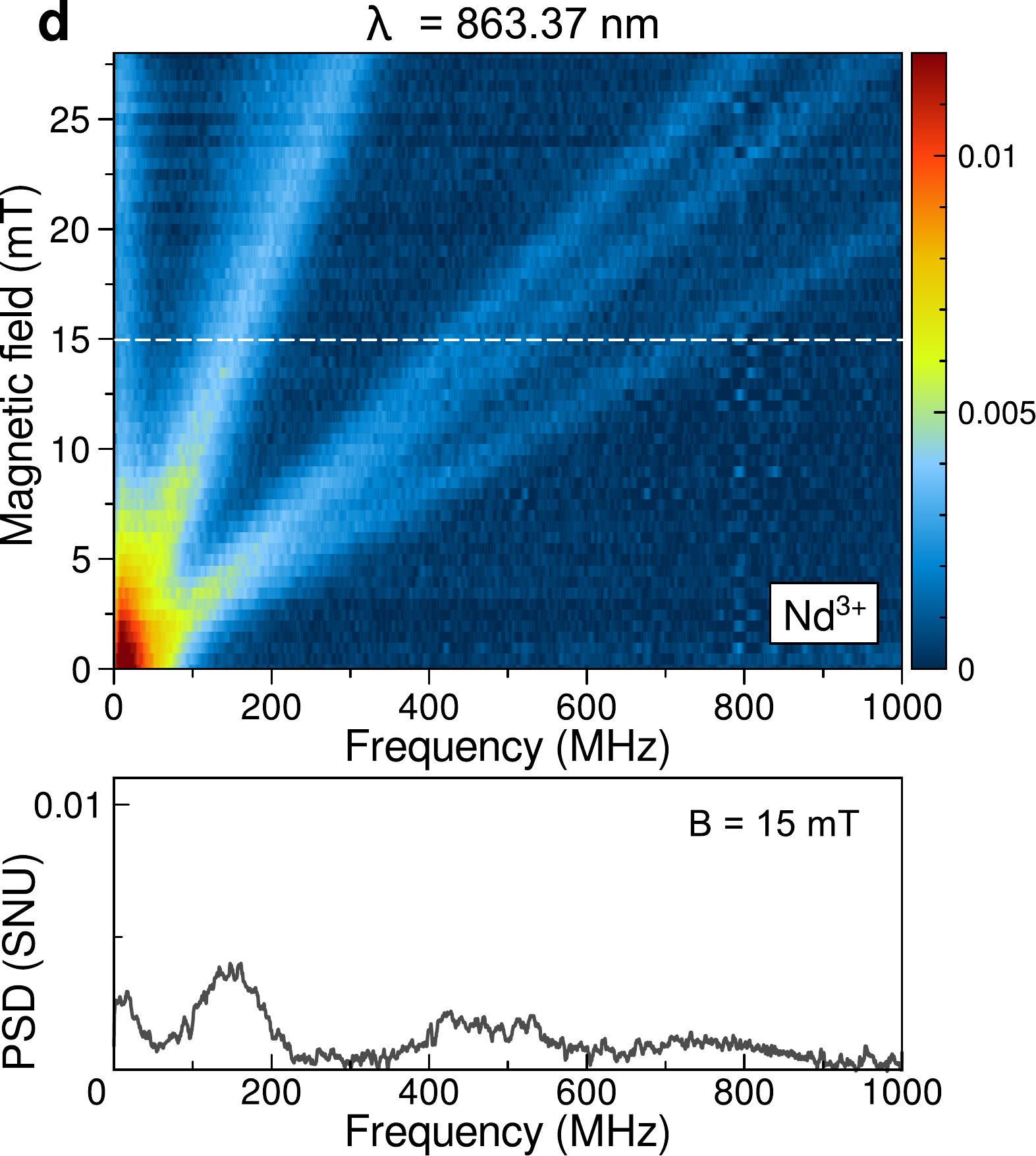}
\includegraphics[width=0.32\linewidth]{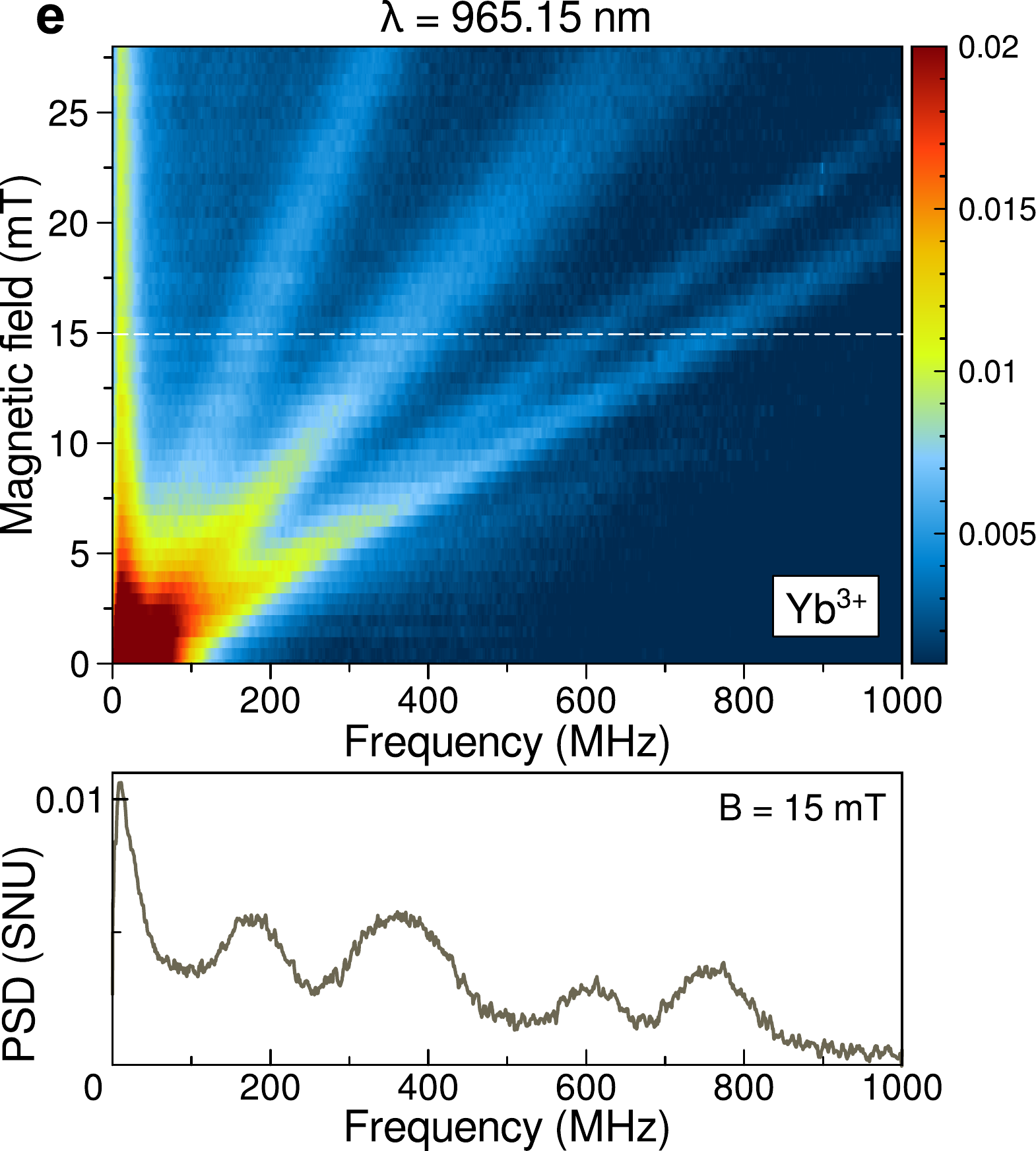}
\caption{\textbf{SN spectra of the SrF$_2$:Nd$^{3+}$(0.5\%)Yb$^{3+}$(0.15\%) crystal.} \textbf{a}, Transmission spectra obtained using a Cary spectrophotometer with a spectral resolution of 0.05\,nm. \textbf{b} and
\textbf{c} High-resolution transmission spectra obtained with a tunable Ti:sapphire laser. Laser power is 1\,mW. The oscillating background signal is produced by laser etaloning at the sample surfaces. \textbf{d} and
\textbf{e} are color maps demonstrating the magnetic field dependence of the SN power spectrum density (PSD) for the Nd$^{3+}$ and Yb$^{3+}$ ions, respectively. The spectra are detected for the same orientation of the
crystal with respect to magnetic field and light propagation. The black lines in the color spectra mark the magnetic fields at which the spectra below in shot noise units (SNU) are taken.
}
\label{fig:2}
\end{figure*}

An important quantity for the SN measurements is the mean-square of fluctuations of the FR
$\langle \delta \varphi^2\rangle$, where $\delta \varphi\equiv\sum_{i=1}^{N'}\varphi_i$. The quantity $\langle \delta \varphi^2\rangle$ can be presented in the form:
\begin{equation}
\langle \delta  \varphi^2\rangle =\langle [\sum_{i=1}^{N'}\varphi_i]^2\rangle  =[\Gamma/\gamma] \varphi_s^2/N .
\label{0}
\end{equation}
In the SNS, we typically detect the spectrum of the noise power or its spectral density, with its characteristic value $W=\langle \delta \varphi^2\rangle T_2$, where $T_2$ is the spin dephasing time. Using Eq.~(\ref{0}),
we can write:
\begin{equation}
	W\equiv \langle\delta \varphi^2\rangle T_2 = {T_2\sigma^2\Gamma\over\gamma}\hskip1mm  {nd\over A}\equiv \eta \hskip1mm  {nd\over A}.
\end{equation}
The parameters entering the last fraction of this equation depend on the measurement conditions and, to a certain extent, can be controlled by the experimentalist. The factor $\eta = T_2\sigma ^2\Gamma/\gamma$ can be
considered as characteristic quantity of the studied paramagnetic center or, better to say, of the particular optical transition addressed. It is this parameter that allows one to compare different spin-systems from the
viewpoint of applicability of SNS. We see that the factor $\Gamma/\gamma$ (which we call the {\it gain factor}) may dramatically enhance the noise power, leaving the FR cross section unchanged.

The above reasoning shows, in particular, that the objects with larger FR cross-section $\sigma$ may be less favorable for the SN technique than those with smaller $\sigma$ but with a larger gain factor $\Gamma/\gamma$.
Figure~\ref{fig:1} shows schematically absorption and FR spectra of paramagnetic RE impurities in crystals in the region of interconfigurational (f-d) and intraconfigurational (f-f) transitions. The FR in the region of
the f-d transitions (easier observable on divalent ions) is usually larger than near the f-f transitions. At the same time, the f-d transition bands with the width $\sim$ 10$^{13}$\,Hz are broadened homogeneously, while
the lines of the f-f transitions with widths around 10$^{10}$\,Hz~\cite{Handbook}, are strongly inhomogeneously broadened with the homogeneous linewidth often in the range of 10$^2$ - 10$^3$\,Hz~\cite{minerals,Mac}, thus
providing spin-noise gain factors of up to 10$^{8}$. It is this effect that allows us to solve the problem of SN detection in activated crystals.\\

\textbf{Experimental results.}
It is clear from the above treatment that the spin-noise gain effect can be observed only when the spectral width of the probe laser light is smaller than the homogeneous linewidth $\gamma$. To meet this requirement, we
used a tunable (cw) ring-cavity Ti:Sapphire laser with a short-term (during 100\,ms) spectral width of $<40$\,kHz. For our study we chose RE ions with the f-f transitions occurring in the spectral range accessible for a
Ti:Sapphire laser (Nd$^{3+}$ and Yb$^{3+}$).

\begin{figure*}[ht]
\centering
\includegraphics[width=\linewidth]{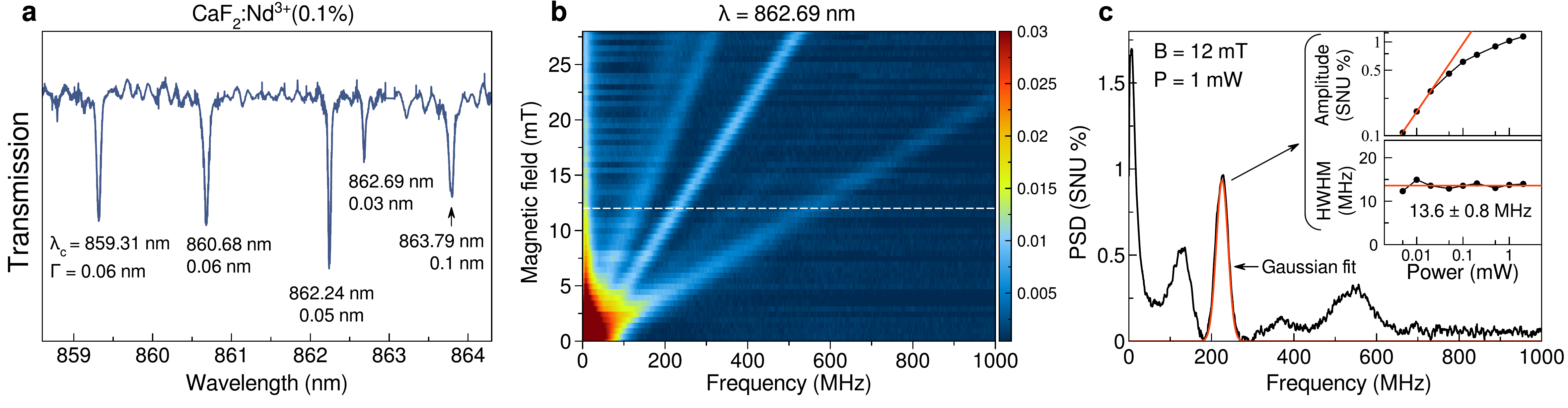}
\caption{\textbf{Spin noise spectra of the CaF$_2$:Nd$^{3+}$(0.1\%) crystal.} \textbf{a}, Laser transmission spectrum of the sample at 5.7\,K. The oscillations are due to the etaloning effect of the laser by the sample.
\textbf{b}, Color map of the SN spectra detected at the absorption peak 862.69\,nm versus magnetic field. Laser power is 1\,mW. \textbf{c}, Exemplary spectrum at $B=12$\,mT with the two insets demonstrating the power
dependence of the amplitude (log-log scale) and the HWHM (semi-log scale) of the peak marked in the main panel.
}
\label{fig:3}
\end{figure*}

The first sample used in our experiments (Sample 1) contained both of these ions - SrF$_2$:Nd$^{3+}$,Yb$^{3+}$. First, overview measurements of transmission spectra of the sample (Fig.~\ref{fig:2}a) were performed to
identify the spectral features for the resonant SN measurements. The most convenient f-f transitions were found in the range 860-870\,nm for Nd$^{3+}$ and 965-980\,nm for Yb$^{3+}$, in agreement with literature
data~\cite{minerals}. For detailed characterization, we performed the same measurements with much higher spectral resolution, using a tunable Ti:sapphire laser (Figs.~\ref{fig:2}b and~\ref{fig:2}c). Narrow spectral features are observed in both wavelength ranges: the Nd$^{3+}$ ion shows one absorption line at 863.37\,nm with the FWHM width $\Gamma=0.08$\,nm obtained using a Gaussian fit; the Yb$^{3+}$ ion demonstrates a strong absorption peak at 965.15\,nm with $\Gamma=0.02$\,nm.

By tuning the laser emission wavelength to the absorption peak of the corresponding lines we detected the strongest SN signal~\cite{OSN}. Figures~\ref{fig:2}d and~\ref{fig:2}e show color maps of the magnetic field
dependencies of the noise power spectra, detected in the Voigt geometry (magnetic field orthogonal to the laser beam direction). Examples of spectra at $B=15$\,mT are given below these maps for the two wavelengths. The
multitude of lines in the SN spectra demonstrates the presence of magnetically non-equivalent anisotropic centers, as typical for a heterovalent substitution of the impurity. Furthermore, the measurements show the
spectral selectivity of the optical SN spectroscopy, especially in comparison with EPR spectroscopy, where the overall signal from the tested sample volume containing different ions is measured.

More careful demonstration experiments were performed using a sample with a single dopant (Sample 2) -- a CaF$_2$:Nd$^{3+}$(0.1 mol.\%) crystal. Figure~\ref{fig:3}a shows the laser transmission spectrum of the sample at
T=5.7\,K. We tested the SN signal at all five absorption peaks observed in this range, and managed to detect strong SN signal only at the 862.69\,nm peak (Fig.~\ref{fig:3}b). In the color map we similarly observe several
lines with different slopes vs. magnetic field. Figure~\ref{fig:3}c demonstrates an exemplary spectrum of the SN power density in shot-noise units at $B=12$\,mT for the probe light power of 1\,mW. Full information about
the symmetry and structure of the centers revealed in the detected SN spectra can be obtained from their orientational dependence, while the field dependence of the resonance linewidth can provide information about spin
relaxation rate in the system.

The method we use here to implement the SN gain effect implies resonant probing of the system and, generally, cannot be considered as non-perturbative. So, it is interesting to study the probe light power dependence of the
magnetic resonance linewidth detected under these conditions. For this purpose we measured the light power dependence of the amplitude and linewidth of the peak at 225\,MHz (see insets in Fig.~\ref{fig:3}c). The amplitude
dependence (plotted in log-log scale) demonstrates an initial linear increase followed by saturation, while the linewidth in this range of light power densities remains constant at 13.6\,MHz. Therefore, we can conclude
that the resonant laser light, probing forbidden electronic transitions of the RE ions, does not affect noticeably their spin dynamics for the used experimental conditions.\\

\textbf{Conclusions.}
In this paper, we have shown that the effect of spin noise gain that should be observed in many paramagnetic impurities with unfilled electronic shells may considerably widen the class of objects of the SN spectroscopy and strengthen potential of the EPR spectroscopy of the impurity paramagnets. The SN technique provides the EPR spectroscopy with additional degrees of freedom inaccessible for its classical version. Thus, the optical channel provides spectral selectivity of impurity centers and high spatial resolution both in transverse and in longitudinal directions (3D tomography). It is highly important that the SN measurements can be performed in a wide frequency range not restricted to the frequency bandpass of photodetectors. EPR measurements in SN spectroscopy are not rigidly coupled to any fixed radio-frequency, as is usually the case in state-of-the-art EPR spectrometers, and allow one to directly obtain panoramic EPR spectra over a wide spectral range. It is also important that the SN measurements do not imply a magnetic polarization of the studied system and, thus, can be performed at high temperatures or in low magnetic fields. We believe that these specific features of the SN spectroscopy will be able to essentially advance the present-day magnetic resonance spectroscopy of crystals and glasses with paramagnetic impurities.\\

\textbf{Methods}\\
\textbf{Sample structure.}
Sample-1 under study is a SrF$_2$ crystal with 0.5\,mol\,\% of Nd$^{3+}$ and 0.15\% of Yb$^{3+}$ ions. The thickness of the sample is 1.85\,mm. Sample-2 is a CaF$_2$:Nd$^{3+}(0.1\%)$ crystal with 0.1\,mol\,\% of Nd$^{3+}$
and 2.15\,mm thickness.

\textbf{Experimental setup.}
To measure the position of the spectral lines of the different rare-earth ions in transmission, we scanned the laser wavelength in the range of expected transitions while modulating its intensity by a chopper at the
frequency of 2\,kHz. The detection behind the sample was done using a single photodiode with lock-in technique. To detect the SN signal, the laser was power- and frequency-stabilized, and transmitted through a single-mode
fiber to symmetrize the spatial mode. We used homodyne detection of the SN~\cite{homodyne1,homodyne2}. For this purpose, the laser was coupled into the entrance port of a Mach-Zehnder interferometer, with one beam going
through the sample and the second one around the sample. The light scattered by the sample has orthogonal polarization to the incident beam and is then collected and combined with the second beam of the interferometer. The resulting interference is measured using balanced photodiodes with 670\,MHz bandwidth. The signal is digitized and fast Fourier transformed by an FPGA~\cite{CrookerFPGA}. The normalized difference between the signal with
transmitted and with blocked scattered light is measured at identical external conditions. The sample is placed in a cold-finger flow-cryostat at the temperature of $T=5.7$\,K. The magnetic field is applied by an
electro-magnet in the Voigt configuration, i.e., orthogonal to the probe beam. The laser is focused on the sample to a spot diameter of 40\,$\mu$m, using a 200\,mm lens.\\

\textbf{Acknowledgments.} We acknowledge financial support by the Deutsche Forschungsgemeinschaft in the frame of the International Collaborative Research Center TRR 160 (Project A5) and the Russian Foundation for Basic Research (Grant No. 19-52-12054). Theoretical part of the work was supported by the Russian Science Foundation (Project 17-12-01124).

\end{document}